# Quantitative analysis of proximity effect

# in Nb/Co$_{60}$Fe$_{40}$, Nb/Ni, and Nb/Cu$_{40}$Ni$_{60}$ bilayers

Jinho Kim, Jun Hyung Kwon, and K. Char*

*Center for Strongly Correlated Materials Research, School of Physics*

*Seoul National University, Seoul 151-742, KOREA*

Hyeonjin Doh

*Department of Physics, Bk21 Physics Research Division and Institute for Basic Science*

*Research, Sung Kyun Kwan University, Suwon 440-746, KOREA*

Han-Yong Choi

*Department of Physics, Bk21 Physics Research Division and Institute for Basic Science*

*Research, Sung Kyun Kwan University, Suwon 440-746, KOREA,*

*and Asia Pacific Center for Theoretical Physics, Pohang 790-784, KOREA*

We have studied the behavior of the superconducting critical temperature $T_c$ in Nb/Co$_{60}$Fe$_{40}$, Nb/Ni, and Nb/Cu$_{40}$Ni$_{60}$ bilayers as a function of the thickness of each ferromagnetic metal layer. The T$_c$'s of three sets of bilayers exhibit non-monotonic behavior as a function of each ferromagnetic metal thickness. Employing the quantitative analysis based on Usadel formalism of the effect of the exchange energy, we observed that the T$_c$ behavior of Nb/Co$_{60}$Fe$_{40}$ bilayers is in good agreement with the theoretical values over the entire range of the data. On the other hand, the T$_c$'s of Nb/Ni and Nb/Cu$_{40}$Ni$_{60}$ bilayers show a higher value in the small thickness regime than the theoretical prediction obtained from the calculation, which matches the dip position and the saturation value of T$_c$ in the large thickness limit. This discrepancy is probably due to the weakened magnetic properties of Ni and Cu$_{40}$Ni$_{60}$ when they are thin. We discuss the values of our fitting parameters and its implication on the validity of the current Usadel formalism of the effect of the exchange energy.

PACS numbers: 74.45.+c, 74.62.-c, 74.62.Yb

*To whom correspondence should be addressed. E-mail: kchar@phya.snu.ac.kr

**Introduction**

The coexistence of superconductivity and magnetic order has been studied for several decades and it is well known that ferromagnetism is detrimental to superconductivity. However, the two orders can coexist under special circumstances. Larkin and Ovchinnikov, and Fulde and Ferrell (LOFF) suggested that in a superconductor containing localized magnetic moments, the superconducting order parameter can survive with a spatial modulation caused by the effect of an exchange field on the Cooper pairs.[1,2] Experimental observations of this state have not yielded success until recently, because the strength of exchange energy must be within a certain narrow range.[3] This experimental difficulty can be avoided when we are dealing with a superconductor (S)/ferromagnetic-metal (F) heterostructure in which the superconductivity is induced in the ferromagnetic region via the proximity effect. In this system, superconductivity can coexist in a thin layer in the ferromagnetic region near the SF boundary. This system offers another way to study the coexistence of superconductivity and magnetism.

The proximity effect between superconductivity and magnetism has been studied for some time. It had been believed that there is a large suppression of the superconducting order parameter at the SF interface due to the strong pair-breaking by the ferromagnet via spin-flip scattering and/or spin-rotation at the interface. This strong suppression of the order parameter had often been taken into account by imposing a vanishing boundary condition at the SF interface. In more recent theoretical studies of the proximity effect in SF heterostructures,[4-8] however, it has been pointed out that the $T_c$ of the heterostructure is expected to exhibit an oscillatory behavior as a function of ferromagnetic metal thickness due to the modulation of the order parameter by the exchange energy, much like the LOFF state. In particular, the $T_c$ oscillation in SF superlattices and in SFS trilayers has been ascribed to the $\pi$-phase coupling of the altering superconducting layers.[4,5]

The $T_c$ behavior in SF heterostructure has been investigated experimentally by many groups. Strunk *et al.* studied a $T_c$ behavior in Nb/Gd/Nb triple layer systems.[9] The authors observed not oscillatory but steplike behavior, which was attributed to the ferromagnetic transition of Gd films of a certain thickness. The $T_c$ oscillation was observed by Jiang *et al.* in Nb/Gd multilayer system and they ascribed the oscillation to $\pi$-phase coupling.[10] Muhge *et al.* observed similar behavior in Fe/Nb/Fe trilayers but they pointed out that the magnetic dead layer plays a dominant role in the oscillation.[11] In Ref. 8 and 12, non-monotonic $T_c$ behavior in Nb/CuNi bilayer systems was reported and analyzed in a quantitative way. An oscillatory $T_c$ behavior in Nb/Ni system and its quantitative analysis was reported in Ref. 13. There have been many experimental efforts to confirm $T_c$ oscillation in many kinds of SF heterostructures besides the works mentioned above.[14-20]

In this paper, we present our experimental results on $T_c$ behavior in SF bilayer systems with Nb as a superconductor and three kinds of ferromagnetic metal, each having a different Curie temperature: Nb/Co$_{60}$Fe$_{40}$, Nb/Ni, and Nb/Cu$_{40}$Ni$_{60}$ bilayers. We also present a quantitative analysis of our data using Usadel formalism, taking into account only the exchange energy inside the ferromagnetic metals as the pair-breaking mechanism.

**Sample fabrication and measurements**

All the samples were deposited with a multi-source DC magnetron sputtering system at ambient temperature using an oxidized Si wafer as the substrate. The thickness of the oxidized layer was 200 nm and the lateral size of the strip-shaped substrates was $2 \times 7$ mm$^2$. After the chamber was evacuated to $2 \times 10^{-8}$ Torr, Nb films were deposited using 99.999 % pure argon gas at 4mTorr. A solid Nb (99.95 % pure) target was used as a source and the deposition rate was 0.29 nm/sec. To minimize run by run scatter in $T_c$ which can be caused by a small difference in the sample preparation conditions, we deposited the Nb layer simultaneously on several samples arranged in a line. The uniformity of the $T_{cS}$ of each Nb sample was measured separately and found to be within 20 mK, and the uniformity of the $T_c$ of each Nb(24 nm)/Co$_{60}$Fe$_{40}$(> 7 nm) sample was found to be within 30 mK. Then, ferromagnetic layers were deposited *in-situ* immediately after Nb deposition, to avoid possible contamination and oxidation of the Nb film surface. To obtain a systematic variation in the thickness of the ferromagnetic layers, we used the natural gradient of the sputtering rate caused when the stage of the substrates is placed in an asymmetric position relative to the center of ferromagnetic target. For the deposition of ferromagnetic layers, 99.9 % pure $Co_{60}Fe_{40}$, 99.98 % pure Ni, and 99.95 % pure $Cu_{40}Ni_{60}$ solid targets were used respectively as sources. The deposition rates of each ferromagnetic layer were 0.1 nm/sec for $Co_{60}Fe_{40}$, 0.13 nm/sec for Ni, and 0.14 nm/sec for $Cu_{40}Ni_{60}$ respectively. As a final step, all the samples were capped with 2 nm of Al to prevent oxidation in the air. The layer thickness was controlled by its deposition time during growth with its deposition rate calibrated using a profilometer. The calibrated thickness of the part of the samples was confirmed by the measurement using transmission electron microscopy and low-angle x-ray diffraction.

The superconducting transition temperatures of the bilayers were measured resistively using a standard DC 4-probe technique with the current magnitudes at 0.1 mA and determined as the temperature at which the resistance of the samples reaches 10 % of the normal state resistance at T = 10 K.

**Results and analysis**

Typical R vs. T curves near $T_c$ normalized to the normal state resistance at T=10 K can be seen in Fig. 1, with varying $d_{Ni}$ for $d_{Nb} = 22.5\,\text{nm}$. All the R vs. T curves during the transitions are almost parallel to each other and exhibit sharp transition. These transition curves correspond to $T_c$'s represented by empty square symbols in Fig. 3(a).

Figure 2 shows the $T_c$ behavior of Nb/Co$_{60}$Fe$_{40}$ bilayers as a function of Co$_{60}$Fe$_{40}$ thickness $d_{CoFe}$ with fixed thickness of Nb, $d_{Nb} = 26$ nm for two different sets of samples. As can be seen in the data, the $T_c$ of the bilayers decreases monotonically from $T_{cS}$ of a single Nb layer with increasing thickness of Co$_{60}$Fe$_{40}$ until it reaches about 2 nm and then increases slightly but definitely to approach a limiting value, resulting in a shallow dip feature of about 100 mK, well above our experimental resolution. We analyzed this data using the method in Refs. 8 and 21 based on Usadel formalism.[22] In this formalism, only the influence of exchange field was included by ignoring the effects of the spin-flip scattering and/or spin-rotation.

To fit our data, we determined the resistivity values of each layer $\rho_{Nb} = 14.6\,\mu\Omega\text{cm}$, $\rho_{CoFe} = 14.8\,\mu\Omega\text{cm}$, and superconducting transition temperature of single Nb layer $T_{cS} = 7.94$ K from separate experiments. Fixing these values, we obtained the following parameters from the fitting: the dirty limit coherence length of superconductor and ferromagnetic metal $\xi_S = \sqrt{\hbar D_S / 2\pi k_B T_{cS}}$ and $\xi_f = \sqrt{\hbar D_f / 2\pi k_B T_{cS}}$, the parameter which represents interface resistance between superconductor and ferromagnetic metal $\gamma_b \equiv \dfrac{R_b A}{\rho_F \xi_F}$, where $R_b A$ is a resistance at SF boundary, and the exchange energy in ferromagnetic metal $E_{ex}$. $E_{ex}$

determines the dip position which is the thickness of the magnetic film where the dip of the $T_c$ vs. $d_{CoFe}$ arises. On the other hand, the parameter which have largest influence on the saturated $T_c$ value is $\gamma_b$. The best result was obtained, yielding $\xi_S \approx 8.3$ nm, $\xi_f \approx 14.4$ nm, $\gamma_b \approx 0.34$, and $E_{ex} \approx 99.4$ meV. The solid line in Fig.2 represents this result. The line exhibits excellent agreement with the data.

We estimated the mean free path of the Nb film from the coherence length, $\xi_S$. From the definition in the dirty limit $\xi_S = \sqrt{\hbar D_S / 2\pi k_B T_{cS}}$ and the diffusion constant $D_S = \frac{1}{3} v_F l_{Nb}$, we obtained the mean free path $l_{Nb} \approx 2.4$ nm when a Fermi velocity of $v_F = 0.56 \times 10^6$ m/sec [23] was inserted. From this value of the mean free path, we obtained $(\rho l)_{Nb} = 3.56 \times 10^{-16}$ $\Omega$m$^2$, which is comparable to the value in Ref. 24. The interface resistance at SF boundary estimated from $\gamma_b$ is to be $R_b A = 0.6 \times 10^{-11}$ $\mu\Omega$cm$^2$, which can be understood as a comparable value to the intrinsic resistance of the interface between two metals.[25] The exchange energy in the LOFF picture is supposed to be the exchange splitting of the conduction band of a ferromagnetic material, which is usually several times higher than the Curie temperature. However, the exchange energy determined from our fit is close to its Curie temperature T$_{Curie}$, about 1200 K, via the relation $E_{ex} = k_B T_{Curie}$, as was found in Ref. 8. Even for the Nb/Ni and Nb/Cu$_{40}$Ni$_{60}$ case, the exchange energies found from the dip position of the $T_c$ data were all very close to their Curie temperatures, well below the theoretically predicted exchange energy. This may point to the fact that the mechanism that breaks the superconducting pairs at the SF interface may be a lot more complicated than the current Usadel picture represents. Except for the lower exchange energy than expected, we conclude that the fitting yields parameters in a reasonable range, and the $T_c$ behavior of the Nb/Co$_{60}$Fe$_{40}$ bilayers is quantitatively consistent with the theory based on Usadel formalism.

The non-monotonic $T_c$ behavior is reproduced when we are using Ni and Cu$_{40}$Ni$_{60}$ as a ferromagnetic metal instead of Co$_{60}$Fe$_{40}$, with a fixed $d_{Nb} = 22.5$ nm although there is some difference in several aspects. Figure 3(a) shows the $T_c$ behavior of Nb/Ni bilayers as a function of Ni thickness for three different series of samples. As can be seen in Fig. 3(a), the $T_c$'s of Nb/Ni bilayers show comparatively slow decrease with increasing Ni thickness $d_{Ni}$ until it reaches ~ 2 nm and then decreases rapidly until the dip position, $d_{Ni}$ ~ 3 nm, is reached. Then, $T_c$'s increase slightly to the limiting value resulting in a dip feature of about 120 mK. The $T_c$ behavior of Nb/Cu$_{40}$Ni$_{60}$ bilayers for two series of samples is shown in Fig. 4(a). The shallow dip feature of about 60 mK is observed where Cu$_{40}$Ni$_{60}$ thickness $d_{CuNi}$ ~ 4 nm and the slow decrease in $T_c$ below the dip position can again be seen in this data.

The $T_c$ behavior of Nb/Ni bilayers as a function of Ni thickness cannot be fitted with fixed parameters over the entire range of ferromagnetic metal thickness, especially where the thicknesses of ferromagnetic metal are small. The same situation arises in Nb/Cu$_{40}$Ni$_{60}$ bilayers as well.

For the analysis, we estimated the superconducting coherence length in the S layer from parallelism to Nb/Co$_{60}$Fe$_{40}$ bilayer case, considering the difference in resistivity of Nb films due to different thicknesses. Fixing this value and the parameters determined from other experiments, $\rho_{Nb} = 15.9\ \mu\Omega\text{cm}$, $\rho_{Ni(CuNi)} = 9.68(24.4)\ \mu\Omega\text{cm}$, and T$_{cS}$ for each case, we calculated the T$_c$ of each bilayer using the method mentioned in the Nb/Co$_{60}$Fe$_{40}$ bilayers case, attempting to find a best result which fits the dip position and saturation value of T$_c$. In the calculation, the values of $\xi_f$, $\gamma_b$, and $E_{ex}$ were adjusted to obtain the best result and each of these results are represented by the solid lines in Fig. 3(a) and Fig. 4(a).

The parameters yielding the lines in Fig. 3(a) and Fig. 4(a) are as follows: $\xi_f = 17.85$ nm, $\gamma_b = 0.7$, and $E_{ex} \approx 51.8$ meV ($T_{\text{Curie}} \approx 600$ K) for Nb/Ni bilayers and $\xi_f = 8.8$ nm, $\gamma_b = 0.57$, and $E_{ex} \approx 14.7$ meV ($T_{\text{Curie}} \approx 170$ K) for Nb/Cu$_{40}$Ni$_{60}$ bilayers respectively. These parameters are summarized in Table I. The corresponding interface resistance at the SF boundaries $R_b A \sim 1.2 \times 10^{-11}\ \mu\Omega\text{cm}^2$, which is almost same in both cases, is about twice as large as that of the Nb/Co$_{60}$Fe$_{40}$ interface. We can compare the values of exchange energy for Co$_{60}$Fe$_{40}$, Ni, and Cu$_{40}$Ni$_{60}$ obtained from the fits. The three exchange energy values are all very consistent with their known Curie temperature values.[26]

In addition, our fitting parameters can be compared with previous results by other groups.[8,13] For Nb/Ni bilayers, the value for superconducting coherence length $\xi_S$ for our bilayers is larger than that obtained for Nb/Ni bilayer in Ref. 13 and the dip position for our case is as twice as larger than that in the same reference. This might be due to the difference in the Nb and Ni film properties. On the other hand, the difference in our $\xi_S$ value from that in Ref. 8 is small and the dip position of our Nb/$Cu_{40}Ni_{60}$ bilayer seems consistent with that of Ref. 8 obtained for Nb/$Cu_{43}Ni_{57}$ bilayers when considering the difference in the Curie temperature of CuNi alloys caused by the relative composition of Cu and Ni.

Although the dip and the saturation values of our $T_c$ data of Nb/Ni and Nb/$Cu_{40}Ni_{60}$ bilayers can be fit, the fitting results show a lower $T_c$ value than the data in both cases when the ferromagnetic metal thickness is small. No combination generates a result which fits the data for the entire range of ferromagnetic thickness. This discrepancy between the data and the fitting results can be explained by the suppression of magnetism for thin ferromagnetic films because lower values for the exchange energy increases the $T_c$ value of SF bilayers. The suppression of magnetism was reported in the references in which the proximity effect in SF multilayers with Nb as the superconductor and Gd, Fe, and Ni as the ferromagnetic metals was investigated.[9-11,15] The reason for the suppression was ascribed to the decrease of the number of nearest neighbors due to the finite size and inhomogeneity of thin films,[9,10,15] and to the alloying effects at the interface.[11] In particular, the superconducting transition temperature and saturation magnetization $M_s$ of Nb/Ni multilayers was studied for a very small ($d_{Ni} \leq 2$ nm) Ni thickness region in Ref. 15. In this reference, the $M_s$ was almost zero when $d_{Ni} \leq 1.2$ nm and the $T_c$ of the multilayers showed a slow decrease, which was similar to that of our Nb/Ni bilayers in the same $d_{Ni}$ range. After this thickness, the $M_s$ from the Ni layers started to increase and the $T_c$ of the multilayers approached to zero. The disappearance of superconductivity in this reference was, however, due to the small thickness of Nb. In our case, because we were dealing with a single ferromagnetic layer instead of multiple layers, the measurement of magnetism in the thin film range was not successful due to insufficient magnetic signal for a SQUID magnetometer. However, considering the similarity in $T_c$ behavior between our data and Ref. 15, the suppression of magnetism is highly likely. Figure 3(b) shows the necessary Curie temperature estimated via the relation $E_{ex} = k_B T_{Curie}$ to fit our data fixing other parameters as in the solid line. Its behavior as a function of Ni thickness is similar to that in Ref. 10 even though a different ferromagnetic material was used.

In our experiment, the roughness of the thin ferromagnetic layers was so small that we did not expect suppression of magnetism due to film inhomogeneity. According to the measurement by using atomic force microscope (AFM), the surface RMS roughness of a single Nb layer was 0.155 nm, and the RMS roughness values of 0.5 nm and 1 nm thick ferromagnetic films on 24 nm Nb films did not increase for all three sets of bilayers as can be seen in Table I.

The alloying at the interface is a possible candidate for the suppression, which will inevitably lead to a large $\gamma_b$. However, because the relatively lower value of $\gamma_b$ derived from the $T_c$ values of each curves at large thickness of ferromagnetic metals indicate little alloying, we suggest a different explanation; structural disorder in ferromagnetic material can lower its magnetism. The difference in crystal structure between Nb and ferromagnetic metals can cause the structural disorder in small thickness regions. The $Co_{60}Fe_{40}$ layer, which is mechanically harder than Cu or Ni, may have less structural disorders than the softer Cu or Ni layers. Smaller $\gamma_b$ values in Nb/$Co_{60}Fe_{40}$ bilayers can be regarded to reflect this assumption.

In addition to the reasons mentioned above, there is another origin for a higher $T_c$ for Nb/$Cu_{40}Ni_{60}$ bilayers in small thickness regions. The Curie temperature of Cu and Ni alloys vary with their relative composition. Ferromagnetism is known to appear when the Ni content is over ~ 46 % and to improve with increasing Ni content.[27] In our Nb/$Cu_{40}Ni_{60}$ bilayers, the Ni content of the $Cu_{40}Ni_{60}$ layer decreases from 60 % to 50 % with decreasing the Ni thickness, as can be seen in Fig. 4(b). This x-ray photoemission spectroscopy data was obtained using ThermoVG (UK) Sigma Probe in ultra high vacuum ($2 \times 10^{-9}$ Torr) condition, with monochromatic Al $K_\alpha$ X-ray source. This compositional change, probably due to better wetting of Cu than Ni on Nb, causes a decrease in exchange energy. And as a result, a decrease in the suppression of $T_c$ leads to the higher $T_c$ value of the bilayer. In Fig. 4(b), the necessary Curie temperature to fit the data is also shown.

In addition, the resistivity of this alloy also varies with the relative composition of Cu and Ni.[28] It has its highest value when the relative composition is ~ 50:50. The flux of Cooper pairs penetrating into $Cu_{40}Ni_{60}$ decreases when the resistivity of the layer increases, resulting in a smaller decrease in $T_c$. The $T_c$ of Nb/$Cu_{40}Ni_{60}$ bilayer samples with the smallest thickness of $Cu_{40}Ni_{60}$ ($d_{CuNi} = 1.13$ nm) cannot be obtained even with zero exchange energy. Thus, we must include the increase of resistivity due to the compositional change.

Although we have completely ignored the spin-flip scattering or spin-rotation caused by strong interaction with the localized magnetic moment at the interface, all our $T_c$ data of three sets of bilayers can be fit reasonably well with a Usadel formalism of the effect of the exchange energy. Furthermore, the reasonable scaling of the dip position with the Curie temperature strengthens the exchange energy based Usadel picture. The only discrepancy with the Usadel model was the smaller exchange energies, about $k_B T_{Curie}$, and the weakened magnetism at the interface for the Ni and the $Cu_{40}N_{60}$ cases. The smaller effective exchange energies we have measured suggest that the energy differences each electron of a superconducting pair experiences when entering the ferromagnetic materials are smaller than the conduction band splitting, although they seem scaling with the Curie temperatures. The exact microscopic origin for the smaller effective exchange energy remains to be understood.

In summary, we have presented the non-monotonic $T_c$ behavior in three different sets of SF bilayers. The nonmonotonic $T_c$ behaviors were observed in several sets of samples even though there were small differences in detailed parameters. The difference between the minimum $T_c$ and the saturation value of $T_c$, ranging from 60 mK to 120 mK depending on the ferromagnetic materials, was certainly larger than our experimental error. From the analysis using the theory based on Usadel formalism of the effect of the exchange energy, we observed a good agreement between our data and the theory except for the lower exchange energy than the theory predicts. In Nb/Ni and $Nb/Cu_{40}Ni_{60}$ bilayers, however, we have found evidences for a thin magnetically weakened layer which is responsible for a slow initial decrease in the $T_c$ of the bilayers, possibly due to structural disorder and/or composition change.

This work is partially supported by KOSEF through CSCMR and by MOST through National Program for Tera-Level Nanodevice.

Figure and Table Captions

FIG. 1. Normalized R(T) curves of Nb(22.5nm)/Ni bilayer samples near $T_c$. The number in parenthesis indicates the thickness of Ni of the corresponding sample. The lines are a guide for eyes. Resistance is normalized by the value in the normal state at T = 10 K. The $T_c$ is determined using a 10 % criterion, and these $T_c$'s correspond to the open square symbol in Fig. 3(a).

FIG. 2. $T_c$ of Nb(26nm)/Co$_{60}$Fe$_{40}$ bilayers as a function of $d_{CoFe}$. The different symbols mean two different sets of data. The solid line is a fit result.

FIG. 3. (a) $T_c$ of Nb(22.5nm)/Ni bilayers as a function of $d_{Ni}$. The different symbols mean three different sets of data. The solid line is a fit result. (b) The Curie temperature necessary to fit our data as a function of $d_{Ni}$.

FIG. 4. (a) $T_c$ of Nb(22.5nm)/Cu$_{40}$Ni$_{60}$ bilayers as a function of $d_{CuNi}$. The different symbols mean two different sets of data. (b) Empty square symbols represent the Curie temperature necessary to fit our data (left axis) and solid circle symbols represent the Ni content in Cu$_{40}$Ni$_{60}$ layers on Nb24nm as a function of $d_{CuNi}$ measured using XPS (right

axis).

TABLE I. Summary of the parameters and the data of the three sets of bilayers.

**Fig. 1**

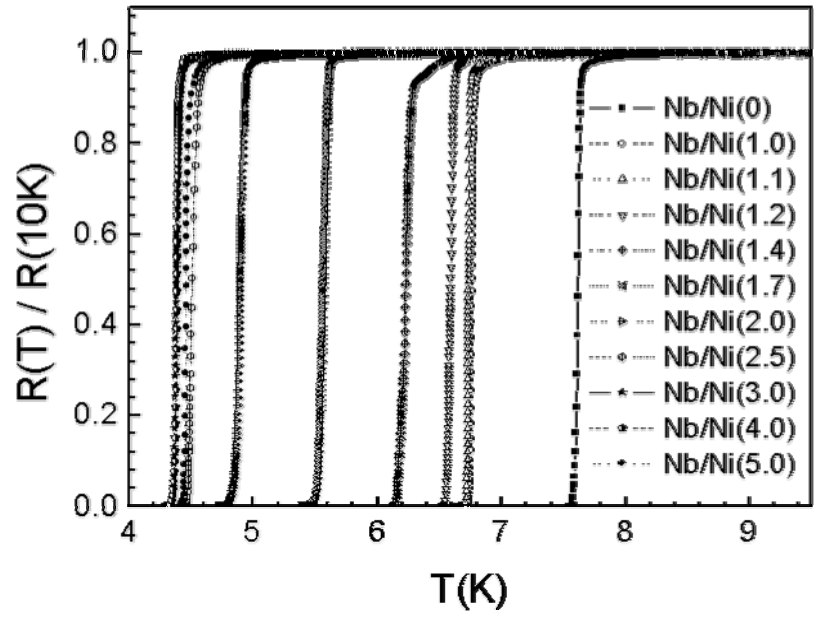

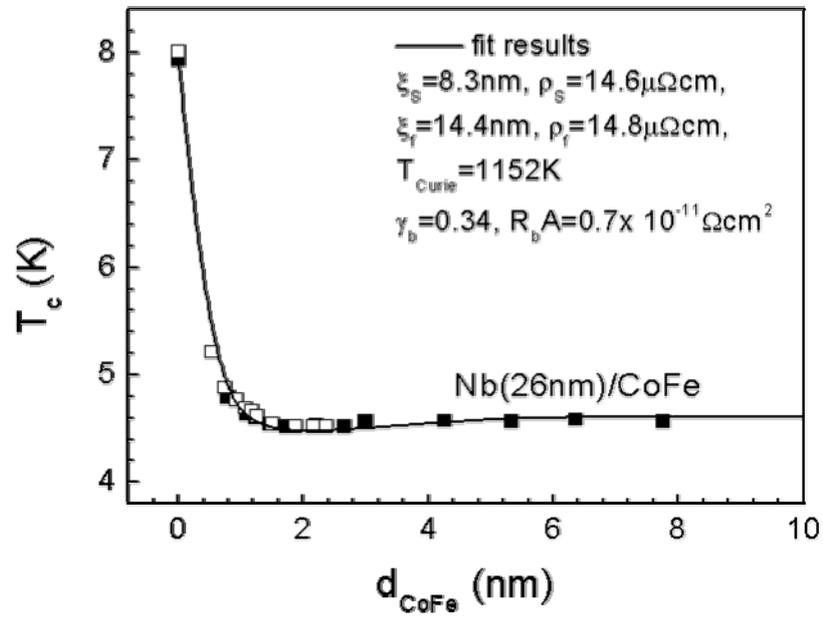

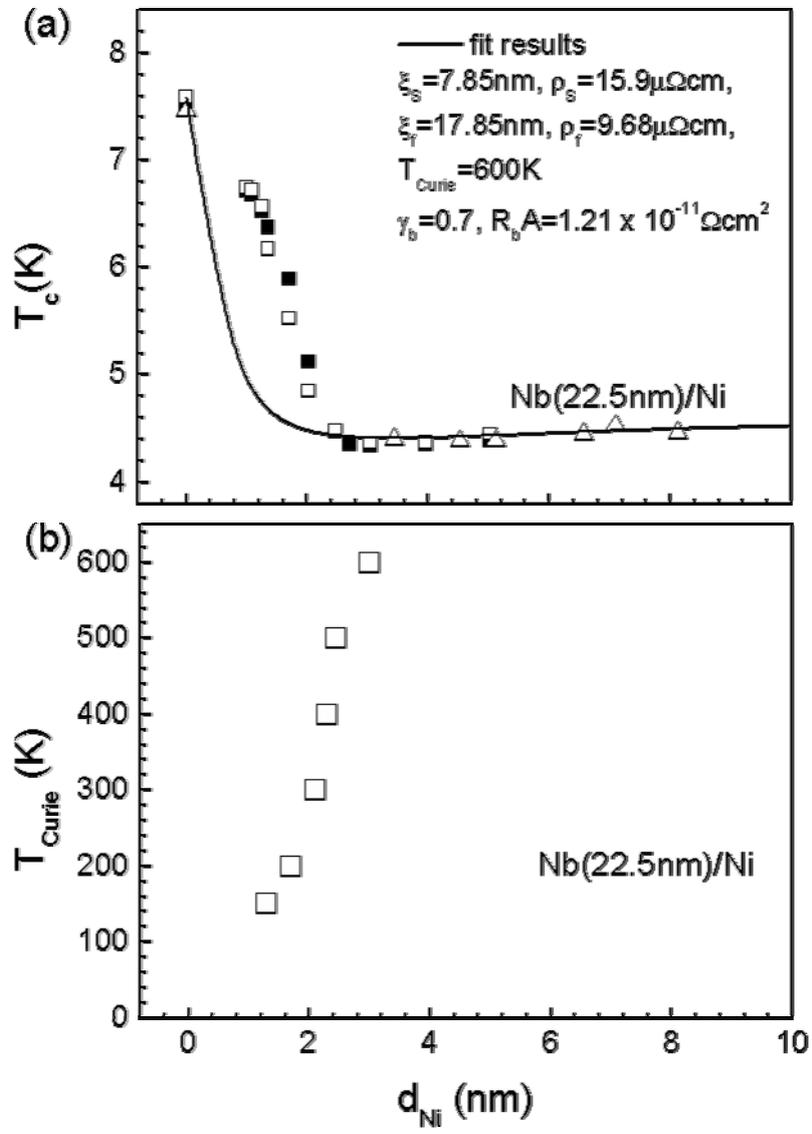

**Fig. 4**

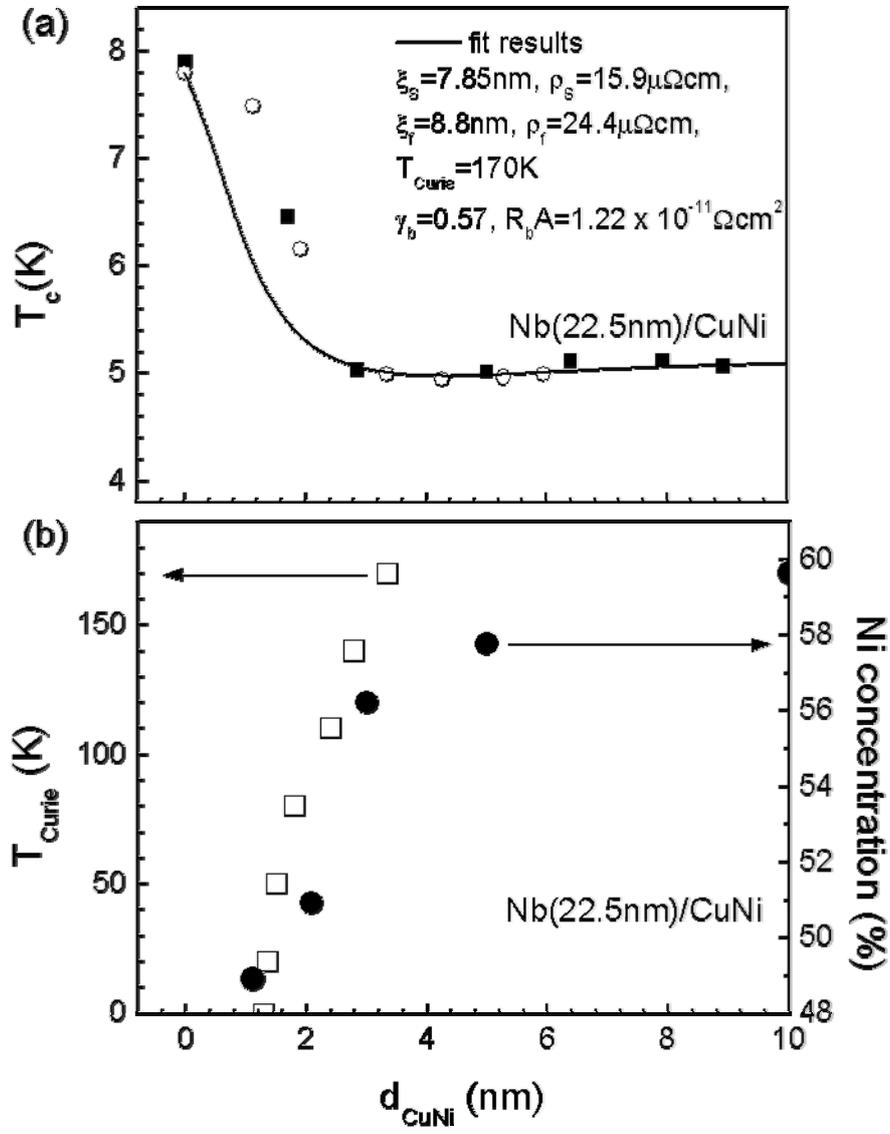

**Table I**

| structure | $\xi_S$ (nm) | $\rho_S$ (μΩcm) | $\rho_f$ (μΩcm) | $\gamma_b$ | $E_{ex}$ (meV) | RMS roughness (nm) | | |
|---|---|---|---|---|---|---|---|---|
| | | | | | | $d_f = 0$ nm | $d_f = 0.5$ nm | $d_f = 1$ nm |
| Nb/Co$_{60}$Fe$_{40}$ | 8.3 | 14.6 | 14.8 | 0.34 | 99.4 | | 0.158 | 0.157 |
| Nb/Ni | 7.85 | 15.9 | 9.68 | 0.7 | 51.8 | 0.155 | 0.166 | 0.161 |
| Nb/Cu$_{40}$Ni$_{60}$ | **7**.85 | 15.9 | 24.4 | 0.57 | 14.7 | | 0.15 | 0.156 |